\documentclass[5p]{elsarticle}

\usepackage[english]{babel}
\usepackage[utf8]{inputenc}
\usepackage{amsmath,amssymb}
\usepackage{mathtools}
\mathtoolsset{showonlyrefs,showmanualtags}
\usepackage{tikz}
\usetikzlibrary{positioning,shapes.geometric}

\usepackage{ctable}
\newcommand{\otoprule}{\midrule[\heavyrulewidth]}
\newcolumntype{C}{>{$}c<{$}}
\newcolumntype{L}{>{$}l<{$}}
\newcolumntype{R}{>{$}r<{$}}
\usepackage{dcolumn}% Align table columns on decimal point
\newcolumntype{d}{D{.}{.}{1}}
\newcolumntype{e}{D{.}{.}{6}}

\newcommand{\coord}[1]{\mathbf{#1}}
\newcommand{\mat}[1]{\boldsymbol{#1}}
\newcommand{\vecr}{\coord{r}}

\newcommand*{\integ}[1]{\int\!\!\!\:\ud{#1}\:}
\newcommand*{\iinteg}[2]{\integ{#1}\!\!\!\integ{#2}}

\newcommand{\crea}[1]{\hat{#1}^{\dagger}}
\newcommand{\anni}[1]{\hat{#1}^{\vphantom{\dagger}}}

\DeclareMathOperator{\Trace}{\mathrm{Tr}}

\DeclarePairedDelimiter{\abs}{\lvert}{\rvert}
\DeclarePairedDelimiter{\norm}{\lVert}{\rVert}
\DeclarePairedDelimiterX\braket[2]{\langle}{\rangle}{#1\delimsize\vert#2}
\DeclarePairedDelimiterX\brakket[3]{\langle}{\rangle}{#1\delimsize\vert#2\delimsize\vert#3}

\newcommand*{\du}{\partial}
\newcommand*{\e}{\textrm{e}}
\newcommand*{\eqspace}{\hphantom{{}={}}}
\newcommand*{\half}{\frac{1}{2}}
\newcommand*{\isDefinedAs}{\coloneqq}
\newcommand*{\Lcal}{\mathcal{L}}
\newcommand*{\Ncal}{\mathcal{N}}
\DeclareMathOperator{\Real}{\mathfrak{Re}}
\newcommand*{\ud}{\mathrm{d}}

\usepackage{hyperref}

\title{Are natural orbitals useful for generating an efficient expansion of the wave function?}

\author[kjhg]{K.J.H.~Giesbertz}
\ead[kjhg]{k.j.h.giesbertz@vu.nl}
\address[kjhg]{Theoretical Chemistry, Faculty of Exact Sciences, VU University, De Boelelaan 1083, 1081 HV Amsterdam, The Netherlands}

\begin{document}

%max 100 words
\begin{abstract}
We investigate whether the natural orbitals (NOs) minimize $\norm{\Psi - \Phi}^2$, where $\Psi$ is a wave function and $\Phi$ is a full configuration interaction (CI) approximation to $\Psi$ in a reduced basis. We will show that the NOs rarely provide the optimal orbitals for $\Phi$, except when (1) there are only two particles or (2) only one basis function is removed in the case of fermions. Further, we will show that the CI expansion coefficients of $\Psi$ and $\Phi$ are identical up to a global scaling factor and demonstrate how the NOs can be used to generate the orbitals that minimize $\norm{\Psi - \Phi}^2$.
\end{abstract}


\maketitle

%%%%%%%%%%%%%%%%%%%%%%%%%%%%%%%%%%%%%%%%%%%%
\section{Introduction}
%%%%%%%%%%%%%%%%%%%%%%%%%%%%%%%%%%%%%%%%%%%%
The claim by Löwdin that the natural orbitals (NOs) are the optimal orbitals for a configuration interaction (CI) expansion of the exact wave function~\cite{Lowdin1955} is generally accepted~\cite{SzaboOstlund1989}. The natural orbitals, $\varphi_k$, are defined as the eigenfunctions of the one-body reduced density matrix (1RDM) defined as
\begin{align}
\gamma(1,1') \isDefinedAs \brakket{\Psi}{\crea{\psi}(1')\anni{\psi}(1)}{\Psi}
= \sum_kn_k\varphi_k(1)\varphi^*_k(1'),
\end{align}
where $1 = \vecr_1\sigma_1$ is a combined space-spin coordinate and the eigenvalues $n_k$ are the natural occupation numbers ($0 \leq n_k \leq 1$) which are assumed to be labelled in decreasing order. Löwdin did not state precisely in which sense the NOs give optimal convergence, but this has been made more precise later by others.
A rather obvious optimality criterion~\cite{KutzelniggSmith1964} is that using the highest occupied NOs, we obtain the best approximation to the 1RDM in $L^2$-norm. In other words, minimizing the distance
\begin{align}
\iinteg{1}{1'}\abs[\Bigg]{\gamma(1,1') - \sum_{i,j}^mb_{ij}f_j(1)g_i^*(1')}^2,
\end{align}
gives $b_{ij} = n_i\delta_{ij}$ and $f_i = g_i = \varphi_i$, the highest occupied NOs. Kobe pointed out that the NOs maximize the contribution of the reference determinant, i.e.\ the NOs ensure that the number of excited determinants is minimal. In other words, the amount of excitations above the pseudo Fermi surface defined by the reference determinant is minimal~\cite{Kobe1969, Davidson1972}. Coleman showed that the NOs are optimal in the sense that expansions of the wave function of the form
\begin{align}
\Phi(1 \dotsc N) = \sum_{i,j=1}^m a_{i,j}\phi_i(1)\theta_j(2 \dotsc N)
\end{align}
are closest to the exact wave function, $\Psi$ in $L^2$-norm, i.e.\ minimizes the distance $\norm{\Psi - \Phi}^2$ if the highest occupied NOs are used for the orbital $\phi_i$. He also showed that the coefficients should be set to $a_{i,j} = \sqrt{n_i}\delta_{i,j}$ possibly multiplied by a phase factor and the functions $\theta_i$ are eigenfunctions of the ($N-1$)RDM~\cite{Coleman1963}, which is exactly the decomposition of the wave function found by Carlson and Keller~\cite{CarlsonKeller1961}.

Unfortunately, these optimality criterions are rarely aimed for in practice. Instead, one would typically aim for a set of orbitals which minimizes the energy or which provides the best approximation to the exact wave function $\Psi$. Since the NOs are not directly related to the Hamiltonian, but reflect the structure of the wave function itself, it is more likely that they are optimal for the expansion of the exact wave function $\Psi$. Such orbitals are of great interest in CI calculations to make the orbital basis as small as possible, to reduce the amount of `dead wood' in the CI expansion~\cite{IvanicRuedenberg2003, BoothThomAlavi2009}. Keeping the size of the orbital space minimal is also very important in the multi-configurational time-dependent Hartree--Fock (MCTDHF) method~\cite{ZhangKollar2013}. The philosophy of MCTDHF is to choose the orbital adaptively while propagating, to reduce the amount of determinants required for a faithful representation of the wave function~\cite{ZanghelliniKitzlerFabian2003, KatoKono2004, CaillatZanghelliniKitzler2005, NestKlamrothSaalfrank2005}.

A decade ago, Bytautas et al.\ have already provided numerical evidence that the NOs actually do not provide the optimal basis for neither of these optimality conditions~\cite{BytautasIvanicRuedenberg2003}: nor for the energy, nor for the normalization deficiency. This latter quantity is calculated by first transforming the CI expansion of the wave function to the basis set of interest. The normalization deficiency is now defined for a subset of determinants, as the deviation of the sum of the squares of their coefficients from unity. The normalization deficiency is therefore a measure how well the exact wave function is approximated by this subset of determinants. Optimal orbitals can now be defined as the ones that minimize the normalization deficiency for a given subset of determinants.

For completeness, we would like to mention that for the purpose of finding orbitals which lead to short CI expansions, the normalization deficiency is not the only suitable measure. A different criterion of recent interest is the seniority number of a CI expansion, which can be interpreted as a measure for the amount of unpaired particles in the expansion~\cite{BytautasHendersonJimenez-Hoyos2011, AlcobaTorreLain2013}. Minimization of the seniority number, therefore, reduces the number of significant determinants in the expansion of the wave function.

The normalization deficiency used by Bytautas et al.\ is equivalent to the normalization criterion under consideration in this letter, $\norm{\Psi - \Phi}^2$, as we will show in more detail later. We will focus on a full CI wave function as an approximation to $\Psi$, constructed from $m$ basis functions
\begin{align}
\Phi(1 \dotsc N) 
= \sum_{\mathclap{i_1\dotsc i_N}}^m d_{i_1\dotsc i_N}\phi_{i_1}(1) \dotsm\phi_{i_N}(N).
\end{align}
The full CI model is generally considered to be too expensive to use in practice. Of course, also other approximate forms of $\Phi$ could be considered. The full CI approximations, however, prevents the mathematical analysis to become overly complicated, while still giving the general idea and leading to some interesting results. Further, the full CI results might be of interest for a recent development based on Monte Carlo techniques to handle up to $10^{29}$ determinants in the full CI space~\cite{BoothThomAlavi2009, ClelandBoothAlavi2010, DadaySmartBooth2012, ClelandBoothOvery2012}. The main idea of this method is to use a set of walkers in the determinant space and to let them evolve according a simple set of rules which include spawning, death and annihilation. Since the walkers corresponding to an insignificant determinant die during the population dynamics, the number of walkers to be handled remains manageable. Finding an optimal basis which reduces the amount of walkers (determinants) required for a faithful representation of the full CI wave function in this technique would make the full CI quantum Monte Carlo (FCIQMC) applicable to even larger systems, since the amount of walkers in the dynamics determines the main computational cost.

In this letter we investigate in more detail why the NOs in general fail to minimize $\norm{\Psi - \Phi}^2$. We do this by first deriving the first order optimality conditions that the optimal orbitals for $\Phi$ need to satisfy to minimize $\norm{\Psi - \Phi}^2$. Next we investigate in which situations the NOs do satisfy these conditions. We will assume that the target wave function $\Psi$ can be expanded in a finite orbital basis of dimension $M$. Note that this assumption does not hold in Coulomb systems, since one would need a complete basis to describe the Coulomb cusp at the coalescence points of the electrons, so $M \to \infty$ in these systems~\cite{Kato1957, GiesbertzLeeuwen2013a, GiesbertzLeeuwen2013b}. The approximate full CI wave function $\Phi$ is constructed out of $m < M$ orbitals of the original orbital basis of the target wave function $\Psi$, so $M - m + 1$ orbitals are eliminated from the original basis set. We will demonstrate that the NOs are rarely the optimal orbitals that minimize the error $\norm{\Psi - \Phi}^2$, confirming the conclusions by Bytautas et al.~\cite{BytautasIvanicRuedenberg2003}. Further, we will investigate to what extend the NOs might still be useful as a sub-optimal choice and how they can be used in procedures to build better orbitals for the approximate wave function that minimize $\norm{\Psi - \Phi}^2$.

%%%%%%%%%%%%%%%%%%%%%%%%%%%%%%%%%%%%%%%%%%%%
\section{First order optimality conditions}
%%%%%%%%%%%%%%%%%%%%%%%%%%%%%%%%%%%%%%%%%%%%
First we will establish the optimal coefficients for the reduced-basis full CI wave function. Since the wave function needs to remain normalized we will work with the following Lagrangian
\begin{align}
\Lcal \isDefinedAs \norm{\Psi - \Phi}^2 - \lambda\Biggl(\sum_{i_1\dotsc i_N}^m\abs{d_{i_1\dotsc i_N}}^2 - 1\Biggr).
\end{align}
For a minimum of $\norm{\Psi - \Phi}^2$ under the normalization constraint, we need that the first order derivatives of the Lagrangian with respect to the expansion coefficients vanish
\begin{multline}
\frac{\du\Lcal}{\du d_{k_1\dotsc k_N}} = \integ{1\dotsm\ud N} \bigl(\Phi^*(1\dotsc N) - \Psi^*(1\dotsc N)\bigr) \\
{} \times \phi_{k_1}(1)\dotsm\phi_{k_N}(N) - \lambda \, d^*_{k_1\dotsc k_N} = 0.
\end{multline}
Now we use the fact that we can make arbitrary transformations of the basis set for our target wave function, so in particular we can choose the first $m$ orbitals to be identical to the (optimal) orbitals for the CI expansion $\Phi$
\begin{align}
\Psi(1 \dotsc N) 
= \sum_{\mathclap{i_1\dotsc i_N}}^M c_{i_1\dotsc i_N}\phi_{i_1}(1) \dotsm\phi_{i_N}(N).
\end{align}
The orbitals $\phi_{m+1}$,\ldots,$\phi_M$ are arbitrary linear combinations of orbitals not contained in the basis set of $\Phi$.

Using the orthonormality of the orbitals, the first order condition from the derivatives with respect to CI coefficients of $\Phi$ can now compactly be written as
\begin{align}
\frac{\du\Lcal}{\du d_{k_1\dotsc k_N}} = d^{\dagger}_{k_N\dotsc k_1}(1-\lambda) - c^{\dagger}_{k_N\dotsc k_1} = 0,
\end{align}
where $d^{\dagger}_{k_N\dotsc k_1} \isDefinedAs d^*_{k_1\dotsc k_N}$. Therefore, we find that the full CI expansion coefficients of the approximate wave function $\Phi$ are identical to the ones of the target wave function $\Psi$ in the optimal basis, up to an overall scaling factor $(1-\lambda)$ to ensure that the approximate wave function $\Phi$ is normalized. The derivatives with respect to $\mat{d}^*$ give the same result, so are not presented here. The same result has been obtained by Zhang and Kollar using a different argumentation~\cite{ZhangKollar2013}.

When searching for the best basis, we should ensure that the basis functions remain orthonormal. Often the orthonormality is enforced by using Lagrange multipliers. An alternative approach is to realize that the allowed variations are in SU($M$), so the unitary variations can also be expressed as~\cite{LinderbergOhrn1977, JensenJorgensenAgren1987, HelgakerJorgensenOlsen2000}
\begin{align}
\mat{U} = \exp(\mat{X}),
\end{align}
where $\mat{X}$ is a traceless anti-hermitian matrix ($\mat{X}^{\dagger} = -\mat{X}$). Since the matrix $\mat{X}$ is anti-hermitian, the unique elements are simply the lower or upper triangle of the matrix and we can perform a free optimization with respect to these unique elements.

Since the reduced-basis CI wave function has the same coefficients as the original wave function up to a global scaling factor for renormalization, the minimization of $\norm{\Psi - \Phi}^2$ is equivalent to the maximization of the norm of $\Phi$ when using the unscaled CI coefficients $\mat{c}$ of $\Psi$
\begin{align}\label{eq:NcalDef}
\Ncal \isDefinedAs \sum_{i_1\dotsc i_N}^m\abs*{\sum_{j_1\dotsc j_N}^Mc_{j_1\dotsc j_N}U_{j_1i_1}\dotsm U_{j_Ni_N}}^2.
\end{align}
This is easily checked, since using $\Ncal = (1-\lambda)^2$ due to the normalization of $\Phi$, we can work out the distance between the wave functions as
\begin{align}
\norm{\Psi - \Phi}^2 = 2 - 2\Real\braket{\Psi}{\Phi} = 2 - 2\sqrt{\Ncal}.
\end{align}
The derivative with respect to the free orbital parameters $\mat{X}$ are readily worked out to be
\begin{align}\label{eq:truncGamCond}
0 = \left.\frac{\du\Ncal}{\du X_{kl}}\right|_{\mat{X}=\mat{0}}
&= \left(\frac{\du}{\du U^{\hphantom{\dagger}}_{kl}} - \frac{\du}{\du U^{\dagger}_{kl}}\right)\Ncal \\
&= \bigl(\theta(m-l) - \theta(m-k)\bigr)\gamma^{(m)}_{kl},
\end{align}
where the Heaviside step function $\theta(x)$ is 1 for positive $x$ and 0 for negative $x$ and we introduced the truncated 1RDM
\begin{align}\label{eq:truncGamDef}
\gamma^{(m)}_{kl} \isDefinedAs 
N\,\sum_{\mathclap{i_2\dotsc i_N}}^m\,c^{\phantom{\dagger}}_{ki_2\dotsc i_N}c^{\dagger}_{i_N\dotsc i_2l}.
\end{align}
Note that the truncated 1RDM is defined for the full original basis set of $\Psi$, so is considered to be an $M \times M$ matrix. Only the summations are reduced from $M$ to $m$.

From the trace of this truncated 1RDM $\Ncal$ can be expressed as
\begin{align}\label{eq:NcalTruncGam}
\Ncal = \frac{1}{N}\Trace\bigl\{\mat{\gamma}^{(m)}\bigr\}.
\end{align}
Hence, if we partition the truncated 1RDM in blocks of orbitals included and excluded from the reduced-basis CI expansion, we find that only the off-diagonal blocks need to be zero. The rotations between the included (excluded) orbitals do not affect $\Ncal$ as expected, since full CI wave functions are invariant with respect to orbital rotations within the CI space.

Naively one would expect that the NOs of $\Psi$ satisfy condition~\eqref{eq:truncGamCond} and that one could simply select the highest occupied NOs to maximize $\Ncal$. However, the NOs of $\Psi$ diagonalize $\mat{\gamma}^{(M)}$ where the sum runs completely up to $M > m$. Since not all CI coefficients of $\Psi$ are included in the sum, the NOs will not diagonalize $\mat{\gamma}^{(m)}$ in general as has already been pointed out by Davidson~\cite{Davidson1972}. There are, however, some special situations in which the NOs of $\Psi$ do satisfy the optimality condition~\eqref{eq:truncGamCond}, so in these cases NOs constitute optimal orbitals for the full CI expansion $\Phi$ in the reduced basis.

\subsection{Two-body systems}
That NOs can provide the optimal orbitals for two-body systems to minimize $\norm{\Psi - \Phi}^2$ is already known for quite some time~\cite{LowdinShull1956, Lowdin1960}. For completeness we will demonstrate explicitly that the NOs of the two-body system indeed satisfy~\eqref{eq:truncGamCond}. The two-body systems are special, since the CI coefficients are only matrices with two indices and the 1RDM becomes a simple matrix product
\begin{align}
\mat{\gamma} = 2\mat{c}\cdot\mat{c}^{\dagger},
\end{align}
so the square root of the natural occupation numbers, $\sqrt{n_k/2}$, are the singular values of $\mat{c}$. In the case of fermions or bosons, the CI coefficients need to be anti-symmetric or symmetric respectively, so restricting to the case of real coefficients, $\mat{c}$ is also normal, hence diagonalizable and the eigenvalues $\xi_k$ are related to the singular values as $\abs{\xi_k} = \sqrt{n_k/2}$.\footnote{The complex case is more involved, since $\mat{c}$ is not normal anymore in general. Symmetric $\mat{c}$ are still diagonalizable~\cite{PhD-Giesbertz2010}. Anti-symmetric $\mat{c}$ can only be block-diagonalized tot $2\times2$ blocks.} Therefore, we find that the NOs block-diagonalize $\mat{c}$ where the blocks only contain degenerate NOs. By diagonalizing the remaining blocks, one obtains a special set of NOs that also completely diagonalizes the CI matrices $\mat{c}$~\cite{Davidson1962}. The truncated 1RDM~\eqref{eq:truncGamDef} in this special NO basis now become
\begin{align}
\gamma^{(m)}_{kl}
= 2\sum_{i=1}^m\,c^{\phantom{\dagger}}_{ki}c^{\dagger}_{il} 
= n_k\delta_{kl},
\end{align}
so are diagonal as well. By selecting the highest occupied NOs one readily maximizes $\Ncal$~\eqref{eq:NcalTruncGam}.

\subsection{Removal of one basis function for fermions}
\label{sec:oneOrb}
In the case of fermions the CI coefficients need to be anti-symmetric to ensure the anti-symmetry of the wave function. We find, therefore, that when only one basis function, $\phi_M$, is removed
\begin{align}
\gamma^{(M-1)}_{Ma}
&= N\,\sum_{\mathclap{i_2\dotsc i_N}}^{M-1}\,c^{\phantom{\dagger}}_{Mi_2\dotsc i_N}c^{\dagger}_{i_N\dotsc i_2a} \\
&= N\,\sum_{\mathclap{i_2\dotsc i_N}}^{M}\,c^{\phantom{\dagger}}_{Mi_2\dotsc i_N}c^{\dagger}_{i_N\dotsc i_2a}
= \gamma^{(M)}_{Ma},
\end{align}
where we used the fact that $c_{MMi_3\dotsc i_N} = 0$ due to the anti-symmetry. Hence, the NOs of $\Psi$ also set $\gamma^{(M-1)}_{Ml} = 0$, so satisfy the stationarity condition~\eqref{eq:truncGamCond}. By simply leaving out the lowest occupied NO, we maximize $\Ncal$~\eqref{eq:NcalTruncGam}. In the case of degeneracy the maximum is not unique and any linear combination of the lowest occupied NOs will do.

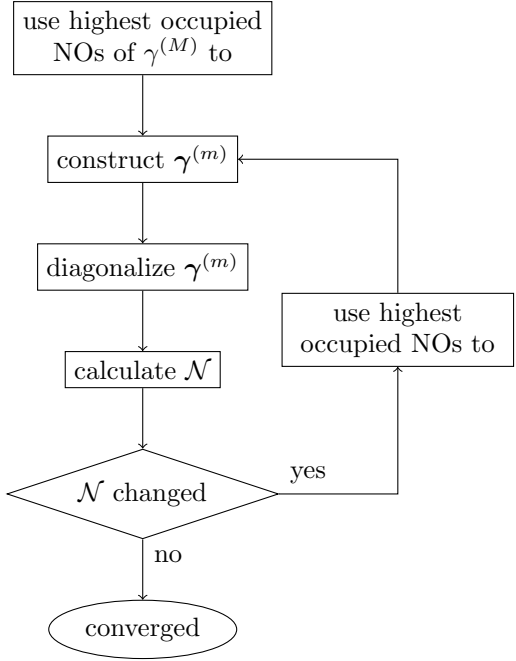
\begin{figure}[t]
  \centering
  \begin{tikzpicture}[node distance=0.8]
    \node (init) [rectangle, draw, text width=90, align=center] {use highest occupied NOs of $\gamma^{(M)}$ to};
    \node (constr) [rectangle, draw, below=of init] {construct $\mat{\gamma}^{(m)}$};
    \node (diag) [rectangle, draw, below=of constr] {diagonalize $\mat{\gamma}^{(m)}$};
    \node (newNcal) [rectangle, draw, below=of diag] {calculate $\Ncal$};
    \node (convTest) [shape aspect=3, diamond, draw, below=of newNcal] {$\Ncal$ changed};
    \node [inner sep=0pt, label=45:yes] at (convTest.east) {};
    \node [inner sep=0pt, label=-45:no] at (convTest.south) {};
    \node (conv) [shape=ellipse, draw, below=of convTest] {converged};
    \node (newNOs) [rectangle, draw, text width=80, right=of newNcal, yshift=15, align=center] {use highest occupied NOs to};
    \path[->]	(init)			edge (constr)
    			(constr) 		edge (diag)
			(diag) 		edge (newNcal)
			(newNcal)		edge (convTest)
			(convTest)		edge (conv);
    \draw [->] (convTest.east) -| (newNOs.south);
    \draw [->] (newNOs.north) |- (constr.east);
  \end{tikzpicture}
  \caption{Primitive optimization of the orbitals for the full CI wave function $\Phi$ in the reduced space.}
  \label{fig:primOpt}
\end{figure}

%%%%%%%%%%%%%%%%%%%%%%%%%%%%%%%%%%%%%%%%%%%%
\section{Calculating optimal orbitals}
%%%%%%%%%%%%%%%%%%%%%%%%%%%%%%%%%%%%%%%%%%%%
We found that the NOs of $\Psi$ are in general not the best orbitals to minimize $\norm{\Psi - \Phi}^2$. To calculate the true optimal orbitals, we need to solve~\eqref{eq:truncGamCond}. The most naive way is to use an iterative scheme where we simply construct a $\mat{\gamma}^{(m)}$ from the NOs of $\Psi$, diagonalize it and take the highest occupied NOs of $\mat{\gamma}^{(m)}$ as new basis functions, reconstruct $\mat{\gamma}^{(m)}$ and iterate this procedure till convergence as depicted in Figure~\ref{fig:primOpt}. This method is identical to a naive implementation of the optimization of the Hartree--Fock (HF) orbitals or Kohn--Sham (KS) orbitals where in each iteration the Fock matrix is diagonalized to the generate new trial orbitals, which are subsequently used to calculate a new Fock matrix till the algorithm converges. This naive implementation is not very efficient to solve for the optimal orbitals for the reduced-basis CI and is even not guaranteed to converge at all. Indeed small test calculations with randomly generated CI matrices for 4 particles in 20 orbitals showed that this iterative algorithm sometimes gets stuck in an alternating solution as also often occurs in HF~\cite{HelgakerJorgensenOlsen2000}.

\begin{figure}[t]
  \centering
  \begin{tikzpicture}[node distance=0.8]
    \node (init) [rectangle, draw] {initialize $i \leftarrow M$};
    \node (diag) [rectangle, draw, below=of init] {diagonalize $\gamma^{(i)}$};
    \node (elim) [rectangle, draw, below=of diag] {eliminate lowest occupied NO};
    \node (iMin) [shape aspect=3, diamond, draw, below=of elim] {set $i \leftarrow i-1$};
    \node [inner sep=0pt, label=45:{$i>m$}] at (iMin.east) {};
    \node [inner sep=0pt, label=-45:{$i = m$}] at (iMin.south) {};
    \node (done) [shape=ellipse, draw, below=of iMin] {done};
    \path[->]	(init)		edge (diag)
    			(diag)	edge (elim)
			(elim)	edge (iMin)
			(iMin)	edge (done);
    \draw [->] (iMin.east) -- +(1.5,0) |- (diag.east);
  \end{tikzpicture}
  \caption{One-by-one elimination scheme to generate an initial guess for the optimal reduced-basis CI orbitals. It is based on the fact that for the truncation of one orbital, the elimination of the lowest occupied NO is optimal (Sec.~\ref{sec:oneOrb}).}
  \label{fig:altInit}
\end{figure}
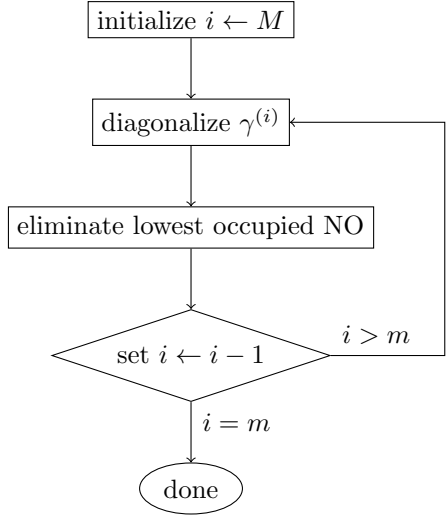

A number of improvements to this naive implementation are obvious. The most important improvement is in the optimization scheme itself, because of its lack of convergence. Since we only need to optimize the orbitals, all the tricks that are available for the optimization of the HF\slash{}KS orbitals can be used to try to achieve and accelerate convergence, e.g.\ the direct inversion in the iterative subspace (DIIS)~\cite{Pulay1980, Pulay1982}. In our case the hessian with respect to the unique elements of $\mat{X}$ can readily be expressed for real $\mat{X}$ as
\begin{align}
\frac{\ud^2\Ncal}{\ud X_{kl}\ud X_{ab}}
= 2\bigl(\Gamma^{(m)}_{kabl} + \Gamma^{(m)}_{kbal} + \delta_{ak}\gamma^{(m)}_{lb} - \delta_{bl}\gamma^{(m)}_{ka}\bigr)
\end{align}
for $a,k \leq m < b,l$ and real upper triangular $\mat{X}$ (see Appendix for more details). The truncated 2RDM is defined as
\begin{align}
\Gamma^{(m)}_{klba} \isDefinedAs 
N(N-1)\,\sum_{\mathclap{i_3\dotsc i_N}}^m\,c^{\phantom{\dagger}}_{kli_3\dotsc i_N}c^{\dagger}_{i_N\dotsc i_3ba}
\end{align}
and comes as an almost free by-product, since the truncated 1RDM has to be constructed anyway from contracting the CI coefficients. Because the hessian is readily available, we have implemented a Newton--Raphson optimization algorithm including the usual trust-region strategy to ensure convergence of the algorithm~\cite{HelgakerJorgensenOlsen2000, NocedalWright1999, ConnGouldToint2000, BazaraaSheraliShetty2006}. This approach is well-known in second-order optimization algorithms for the multi-configuration self-consistent field (MCSCF) wave function~\cite{LinderbergOhrn1977, JensenJorgensenAgren1987, HelgakerJorgensenOlsen2000}. Of course, also other optimization schemes used in MCSCF calculations could be used as well~\cite{IvanicRuedenberg2003, RoosTaylorSiegbahn1980, RuedenbergSchmidtGilbert1982, SchmidtGordon1998} or the iterative approach proposed by Zhang and Kollar~\cite{ZhangKollar2013}. However, the second order method is the most robust one, hence most reliable to provide optimal orbitals.

Now let us consider the initial guess for the reduced-basis CI orbitals. We have seen that the elimination of the lowest occupied NO gives the optimal reduced-basis CI orbitals if only one orbital had to be eliminated. An obvious scheme to generate the initial guess is to continue this procedure to eliminate orbitals one by one (see Figure~\ref{fig:altInit}). Unfortunately, this scheme does not yield directly the optimal orbitals, except under special circumstances. To show this, consider the removal of two orbitals. Since $\phi_{M-1}$ by construction already satisfies its optimality condition $\gamma^{(M-2)}_{M-1a} = 0$ for $a \leq M-2$, we only need to consider
\begin{align}
\gamma^{(M-2)}_{Ma} 
= - N(N-1)\,\sum_{\mathclap{i_3\dotsc i_N}}^M\,c_{MM-1i_3\dotsc i_N}c^{\dagger}_{i_N\dotsc i_3M-1a},
\end{align}
where we used the fact that $\gamma^{(M)}_{Ma} = 0$. The sum on the right-hand side typically only vanishes if $c^{\phantom{\dagger}}_{MM-1i_3\dotsc i_N} = 0$ for some reason. The most obvious case is that $\phi_M$ is a NO with zero occupation number, which means that this NO does not contribute to any determinant. This situation is typical for non-interacting systems, which usually only require one determinant. All unoccupied NOs can straightforwardly all be eliminated. An other possible situation is that the orbitals $\phi_M$ and $\phi_{M-1}$ never co-occur in any determinant contributing to the CI. Both situations do not seem to be likely for fully interacting systems~\cite{GiesbertzLeeuwen2013a, GiesbertzLeeuwen2013b}.

\begin{figure}[t]
  \includegraphics[width=\columnwidth]{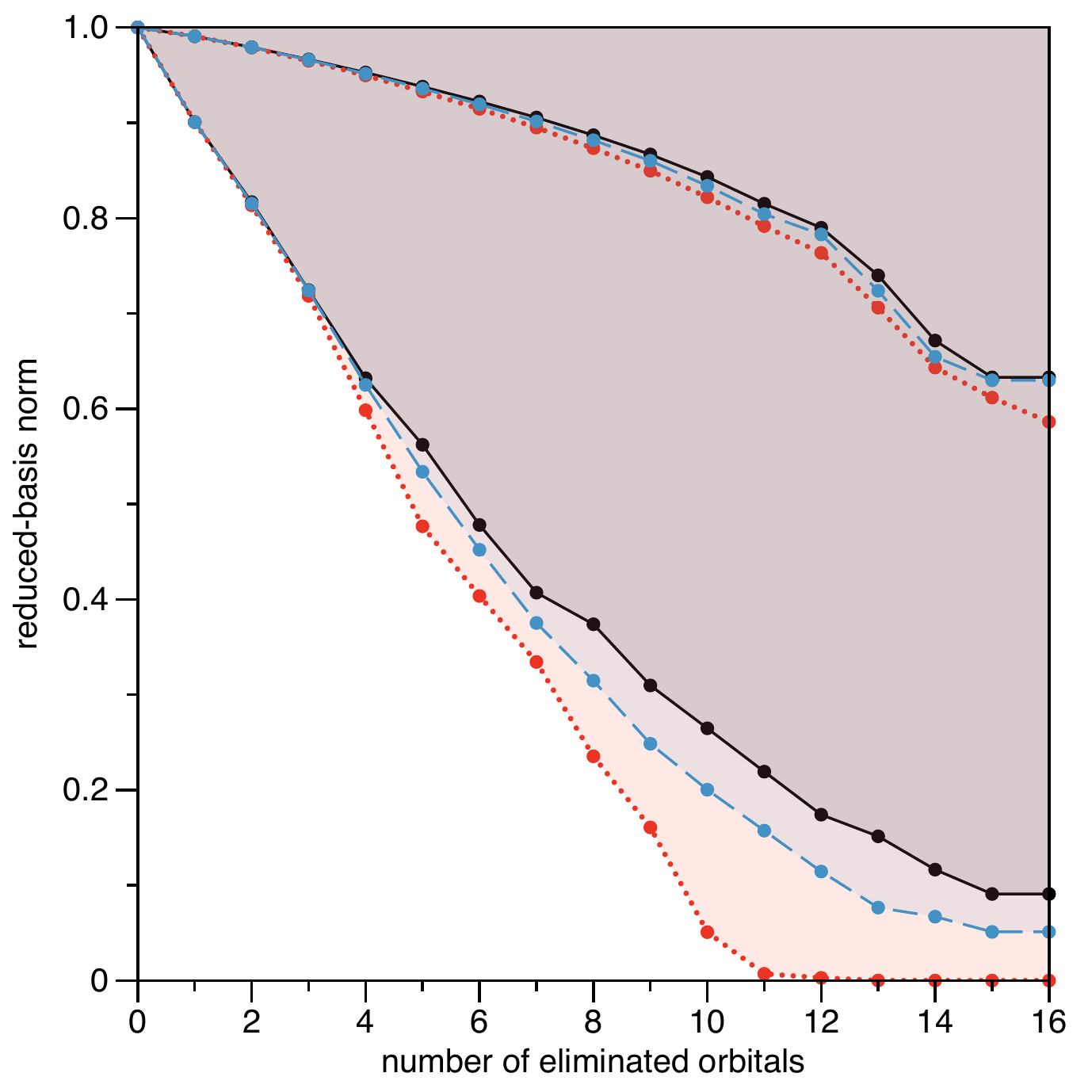}
  \caption{The average (upper lines) and minimum values (lower lines) of $\Ncal$ for an initial basis set of 20 orbitals. The solid lines correspond to the fully optimized orbitals, the dotted lines to the highest occupied NOs and the dashed lines to the one-by-one removal of one NO (Figure~\ref{fig:altInit}). The spread of $\Ncal$ is indicated by the shaded areas and always extends fully up to $\Ncal = 1$, since single determinant wave functions are also present in the test sample.}
  \label{fig:rand}
\end{figure}

Although this alternative guess for the orbitals does not give the best orbitals for the reduced-basis CI, it is expected to give better initial orbitals than just the highest occupied NOs. To check this statement, we have implemented both procedures and the Newton--Raphson procedure to find the optimal orbitals for four fermion CI wave functions. The Newton--Raphson procedure terminated when the norm in the gradient was smaller than $\sqrt{\epsilon} \approx 1.5\cdot 10^{-8}$, where $\epsilon$ denotes the machine precision. The function $\Ncal(\mat{X})$ can have multiple local maxima. To increase the chance that the Newton--Raphson procedure terminates at the global maximum, we run the calculation two times. Once starting from the highest occupied NOs and once from the orbitals generated by the one-by-one elimination scheme.

To generate CI coefficients we used a random number generator and have set the unique CI coefficients to
\begin{align}
c_{i_1i_2i_3i_4} = \frac{\text{ran}_1 - \text{ran}_2}{\text{ran}_3 - \text{ran}_4},
\end{align}
where $\text{ran}_i$ are subsequently generated random numbers. Only using one random number, $c_{i_1i_2i_3i_4} = \text{ran}_1$ did not give sufficient variation in the coefficients to give a good sampling over all possible full CI wave functions. Since the full CI wave functions are completely random, they are not necessarily eigenfunctions of a physical Hamiltonian with only one- and two-body operators.

For each randomly generated CI wave function we have calculated $\mathcal{N}$ for both initial guesses and also its value after full optimization by the Newton-Raphson procedure. In Figure~\ref{fig:rand} we show statistical results for the sampling over $10\,000$ randomly generated CI wave functions for 20 basis functions as a function of the number of eliminated orbitals. The shaded areas indicate the spread of the value of $\Ncal$ for each set of orbitals and the lowest value found, is emphasized by a line [fully optimized (solid), highest NOs (dotted) and one-by-one elimination (dashed)]. The spread completely stretches to $\Ncal =1$ even when the maximum number of orbitals is eliminated. This indicates that our sampling also generated full CI wave functions which can (almost) exactly be represented by one-determinant, which are eigenstates of a non-interacting system of fermions. The fact that we randomly generated one-determinant wave functions gave us confidence that the variation in our sample is sufficient. The lines in the middle of the shaded areas indicate the average values of $\Ncal$ for the different methods.

The smallest value for $\Ncal$ when using the highest occupied NOs (lowest red dashed line in Figure~\ref{fig:rand}) become exponentially small when approaching the maximum of removable orbitals (16). This result is orthogonal to the general assumption that the determinant constructed from the highest occupied NOs always has a significant overlap with the full CI wave function~\cite{AbramsSherrill2004}. However, the full CI wave functions in these worst case scenarios are highly correlated and most NOs have a significant occupation. This large spread of occupation allows the main contribution to the wave function, to come from determinants which only contain one or two of the 4 highest occupied NOs. More details about the worst case for only retaining one determinant can be found in the Supplementary Material. It is not clear if such CI wave functions occur in physical systems. However, a large number of significantly occupied NOs can occur in fractional quantum Hall droplets~\cite{ToloHarju2010}, so this worst case scenario for the NOs might be relevant for these very strongly correlated systems.

\begin{figure}[t]
  \includegraphics[width=\columnwidth]{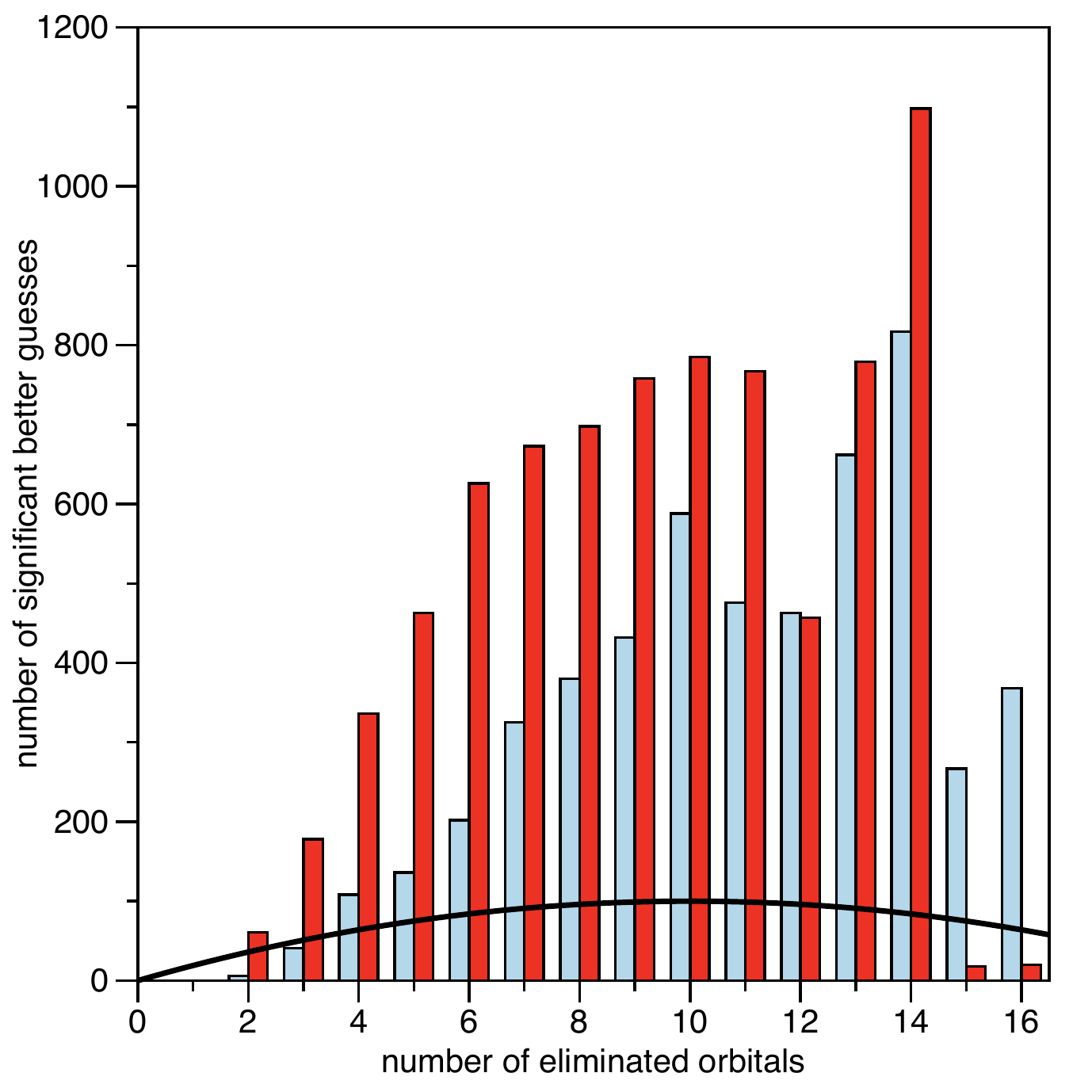}
  \caption{The number of times the initial guess from the NOs (red, right bar) and the one-by-one elimination (blue, left bar) led to significant ($10^{-6}$ times best $\Ncal$) better values of $\Ncal$ after the Newton--Raphson optimization. The black line indicates the number of variational variables in the optimization procedure.}
  \label{fig:initBest}
\end{figure}

An interesting feature of the of plot in Figure~\ref{fig:rand} is that both for the one-by-one elimination and the fully optimized orbitals, there is no difference in the elimination of 15 or 16 orbitals. This is actually a general feature that the reduction to $N$ and $N+1$ orbitals yield the same reduced-basis norm $\Ncal$. For the fully optimized orbitals this can be easily understood. First consider the case of the reduction to $N$ orbitals. Since all orbitals need to be fully occupied, only one Slater determinant can be constructed and the only degrees of freedom are the orbital rotations. When we extend the basis with one additional orbital, we also can have single excited determinants. These additional single excited determinants, however, do not add additional variational freedom to the wave function, so are redundant~\cite{HelgakerJorgensenOlsen2000}. This situation is similar to the addition of determinants to the HF wave function to lower the energy. Adding only single excited determinants does not lower the energy. One needs to add at least double excited determinants to gain in energy, which is closely related to Brillouin's theorem. More explicit proofs can be found in~\cite{ZhangKollar2013} and references therein.

In order to prove this for the one-by-one elimination, more work is required. By definition, we have for the elimination of the last orbital $\phi_{N+1}$
\begin{align}\label{eq:one-by-oneCondN}
0 = \gamma^{(N)}_{N+1,a} 
&= N\,\sum_{\mathclap{i_2\dotsc i_N}}^N\,c^{\phantom{\dagger}}_{N+1i_2\dotsc i_N}c^{\dagger}_{i_N\dotsc i_2a} \\
&= N!\, c^{\phantom{\dagger}}_{N+1\bar{a}_2\dotsc \bar{a}_N}c^{\dagger}_{\bar{a}_N\dotsc \bar{a}_2a}, \quad
\forall \; a=1,\dotsc N,
\end{align}
where $\bar{a}_2 \neq \bar{a}_3 \neq \dotsb \neq \bar{a}_N \in \{1,\dotsc, N\} \backslash a$, so $\bar{a}_2\dotsc \bar{a}_N$ can be seen as a unique complement to $a$. Since we have only one Slater determinant, we necessarily have $c^{\dagger}_{\bar{a}_N\dotsc \bar{a}_2a} = 1/\sqrt{N!}$, so Eq.~\eqref{eq:one-by-oneCondN} implies that $c_{N+1\bar{a}_2\dotsc \bar{a}_N} = 0$. Thus, for the reduction in $\Ncal$, we find
\begin{align}
\Delta\Ncal = N\sum_{\mathclap{i_2\dotsc i_N}}^N\abs{c_{i_2\dotsc i_NN+1}}^2 = 0,
\end{align}
so there is no further reduction in the reduced-basis norm when going from $N+1$ to $N$ basis functions.

Statistically the one-by-one elimination is a clear winner. Both the average value for $\Ncal$ and the worst value of $\Ncal$ are consistently higher if more than one orbital is removed. The methods are identical for the elimination of zero or only one orbital, so for these cases there is no difference and both are optimal. However, in some cases using the highest occupied NOs as an initial guess provide a superior initial guess compared to the one-by-one elimination. Subsequent optimization with the Newton--Raphson procedure starting from the one-by-one could not always overcome this difference. In Figure~\ref{fig:initBest} we have plotted the number of times one of the initial guesses gave a significantly ($10^{-6}$ times best $\Ncal$) more successful result after optimization with the Newton--Raphson procedure. Since the local hessian is readily available, we have checked whether it was negative definite and it always was. The higher solutions found from the less successful initial guess therefore correspond to lower lying local maxima. The existence of multiple local maxima is no surprise, since the exponential Ansatz for the orthogonal variations makes $\Ncal(\mat{X})$ a highly non-linear function.

\begin{figure}
  \includegraphics[width=\columnwidth]{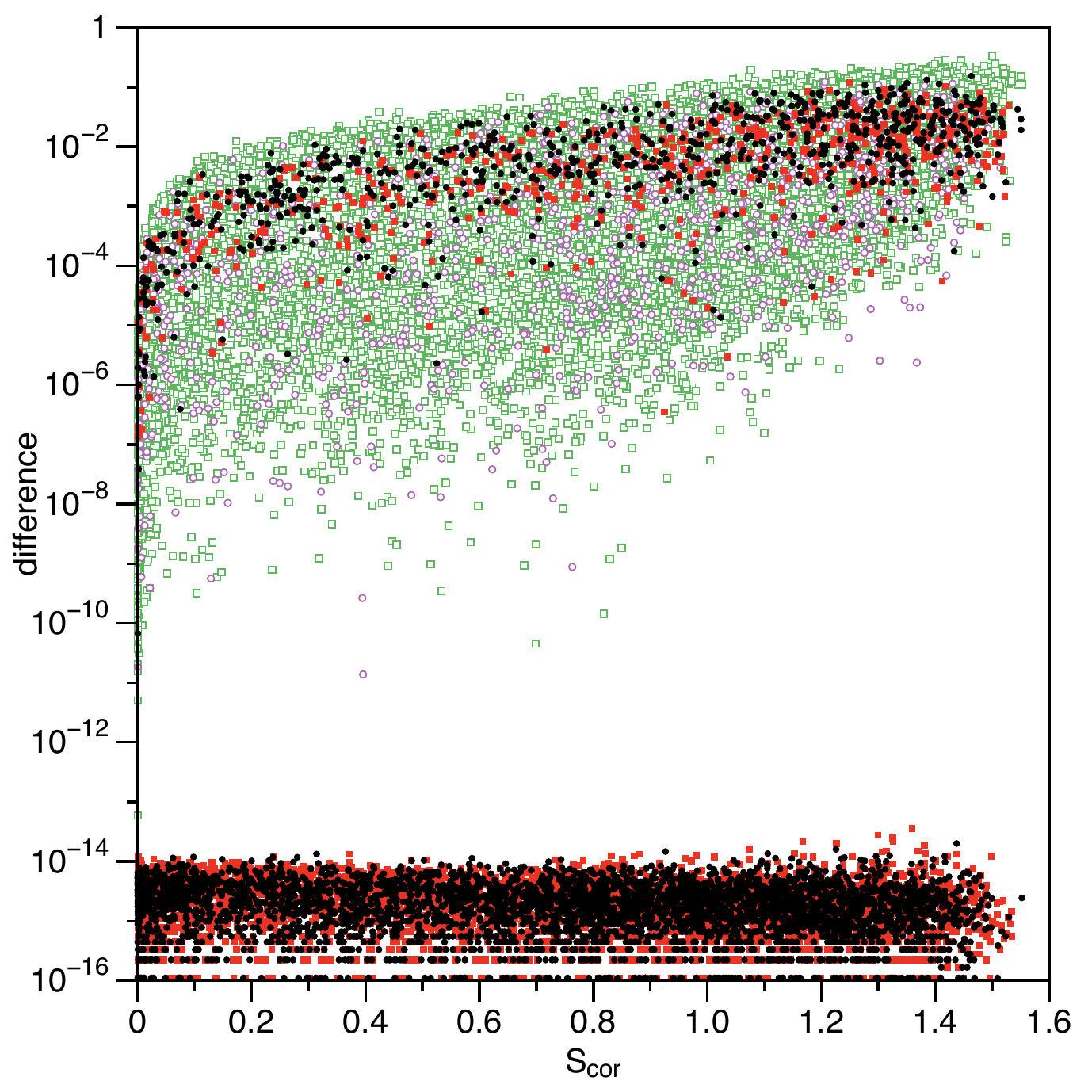}
  \caption{The difference between the two initial guesses as a function of the entropy for the removal of 10 orbitals out of 20 for $10\,000$ randomly generated full CIs. The difference in the initial value for $\Ncal$ is shown by open symbols: (round,purple) for superior NOs and (square,green) for superior one-by-one elimination. The difference in the final values of $\Ncal$ is shown by solid symbols: (round,black) for superior NOs and (square,red) for superior one-by-one.}
  \label{fig:entropy}
\end{figure}

A striking result from Figure~\ref{fig:initBest} is that although the initial value for $\Ncal$ of the NOs is usually inferior to the value form the one-by-one elimination, the NOs provide more often a superior initial guess for the full optimization. Only for the elimination of 15 and 16 orbitals the one-by-one elimination provided significantly more often a better initial guess for the Newton--Raphson algorithm. The dependency of the fully optimized value of $\Ncal$ on the initial guess seems initially to be correlated with the number of degrees of freedom in the optimization, which is indicated by the black curve in Figure~\ref{fig:initBest}. From 12 orbitals onwards, this trent does not seem to hold anymore. Probably, the variational landscape $\Ncal(\mat{X})$ becomes more bumpy, so the optimization becomes more dependent on the initial guess, although the number of variational parameters goes down. Since we have only used two different initial guesses, it is quite likely that the Newton--Raphson procedure did not even find the global maximum in all cases, especially when a large number of orbitals are removed. Possible improvements may be sought in starting from a larger number of initial guesses, which could be generated randomly by making first a random transformation of the basis set as has also been proposed by Zhang and Kollar~\cite{ZhangKollar2013}. An alternative solution would be to use a global optimization scheme like simulated annealing.

It is not so clear in which cases one of the initial guesses is superior over the other. One might expect that there would be a connection with the amount of correlation in the system, which can be quantified as a correlation entropy~\cite{Collins1993, Ziesche2000} defined as
\begin{align}
S_{\text{cor}} \isDefinedAs -\frac{1}{N}\Trace\bigl\{\mat{\gamma}\ln\mat{\gamma}\bigr\} = -\frac{1}{N}\sum_kn_k\ln n_k.
\end{align}
In Figure~\ref{fig:entropy} we show the difference between the initial values of $\Ncal$ of the two initial guesses with open symbols as a function of correlation entropy. The only correlation with respect to the $S_{\text{cor}}$ seems to be the magnitude of the difference. When the initial guesses are optimized by the Newton--Raphson algorithm, these differences are mostly reduced to numerical precision. Only a few significant differences remain (see also Figure~\ref{fig:initBest}). With increasing correlation entropy, both the amount of significant differences and the magnitude of the difference in $\Ncal$ itself increase somewhat. However, both initial guesses perform about equally for any value of the correlation entropy, so the amount of correlation is irrelevant for the choice between the two initial guesses.

%%%%%%%%%%%%%%%%%%%%%%%%%%%%%%%%%%%%%%%%%%%%
\section{Conclusion}
%%%%%%%%%%%%%%%%%%%%%%%%%%%%%%%%%%%%%%%%%%%%
In this letter we have investigated which orbitals minimize $\norm{\Psi - \Phi}^2$, where $\Psi$ is a CI wave function which is approximated by the full CI wave function $\Phi$ in a reduced orbital space. First order optimality conditions revealed that the CI coefficients of both wave functions $\Psi$ and $\Phi$ are identical up to an overall scaling factor which is required to have both wave functions normalized. We have used this information to simplify the optimization problem by reformulating it as a maximization of the norm of $\Phi$ when using unscaled CI coefficients of $\Psi$, $\Ncal$~\eqref{eq:NcalDef}. Setting the first order derivative of $\Ncal$ to zero provided the first order optimality condition that the matrix elements of the truncated 1RDM~\eqref{eq:truncGamDef} between the included and excluded orbitals have to vanish~\eqref{eq:truncGamCond}. This requirement is usually not satisfied by the NOs except when only one orbital is removed in the case of fermions or when dealing with two-electron systems. In the case of two-electron systems these NOs should not only diagonalize the 1RDM, but also $\Psi$ itself~\cite{Davidson1962}.

The similarity of the first order stationarity conditions in terms of the truncated 1RDM and the Fock matrix immediately implies that all the standard solution methods known in HF and DFT to solve their orbital equations to self-consistency can be used. Since the hessian is readily available, we have used the very powerful Newton--Raphson procedure starting from two different initial guess\-es for the optimal orbitals: (1) the NOs and (2) a one-by-one elimination scheme inspired by the fact that the NOs are optimal for the removal of one orbital (Figure~\ref{fig:altInit}). We have run a trial over $10\,000$ CI randomly generated CI wave functions for 4 fermions and 20 basis functions. The one-by-one elimination gave the best initial value for $\Ncal$. The NOs, however, provided more often a better starting guess for the optimal orbitals when a Newton--Raphson procedure was subsequently used to maximize $\Ncal$. We have not been able to find a clear criterion to determine
when one of the initial guesses performs better than the other. Since there is no clear best initial guess, it will be best to use both to generate the optimal reduced-basis CI orbitals in practice.

\section*{Acknowledgements}
The author thanks E.J. Baerends for useful discussions and the Netherlands Foundation for Research NWO for a VENI grant (722.012.013).

\appendix

\section{Hessian}
In this appendix we digress on the calculation of the hessian of $\Ncal$ with respect to the unique entries of $\mat{X}$. First we expand $\mat{U}$ till second order in $\mat{X}$.
\begin{align}
U_{kl} = \bigl(\e^{\mat{X}}\bigr)_{kl} = \delta_{kl} + X_{kl} + \half\bigl(\mat{X}^2\bigr)_{kl} + \dotsb,
\end{align}
so for the first and second order derivatives with respect to $\mat{X}$ we find
\begin{align}
\left.\frac{\du U_{rs}}{\du X_{kl}}\right|_{\mat{X} = \mat{0}} &= \delta_{kr}\delta_{sl}, \\
\left.\frac{\du^2 U_{rs}}{\du X_{kl}\du X_{ab}}\right|_{\mat{X} = \mat{0}}  &= \half\bigl(\delta_{kr}\delta_{sb}\delta_{la} + \delta_{ar}\delta_{sl}\delta_{ka}\bigr)
\end{align}
Now we will use these expressions to calculate the gradient and hessian of $\Ncal$ with respect to $\mat{X}$. First we write $\Ncal$ explicitly in terms of $\mat{X}$ as
\begin{multline}
\Ncal = \sum_{i_1\dotsc i_N}^m\sum_{\substack{r_1\dotsc r_N \\ s_1 \dotsc s_N}}
c^{\hphantom{\dagger}}_{r_1\dotsc r_N}\bigl(e^{\mat{X}}\bigr)_{r_1i_1}\dotsb\bigl(e^{\mat{X}}\bigr)_{r_Ni_N} \\
{} \times \bigl(e^{-\mat{X}}\bigr)_{r_Ni_N}\dotsb\bigl(e^{-\mat{X}}\bigr)_{r_1i_1}c^{\dagger}_{s_N\dotsc s_1}
\end{multline}
The first order derivative with respect to $\mat{X}$ gives~\eqref{eq:truncGamCond} and will not be repeated here. The second order derivative is more involved. Not only we have second order derivatives of one particular orbital rotation, but we also obtain cross-terms
\begin{multline}
\left.\frac{\du^2\Ncal}{\du X_{kl}\du X_{ab}}\right|_{\mat{X} = \mat{0}} \\
\begin{aligned}
&=\Gamma^{(m)}_{aklb}(b,l \leq m) + \half\gamma^{(m)}_{kb}\delta_{al}(b \leq m) + {} \\
&\eqspace \half\gamma^{(m)}_{al}\delta_{bk}(l \leq m) + \Gamma^{(m)}_{aklb}(a,k \leq m) + {} \\
&\eqspace \half\gamma^{(m)}_{kb}\delta_{al}(k \leq m) + \half\gamma^{(m)}_{al}\delta_{bk}(a \leq m) - {} \\
&\eqspace \Gamma^{(m)}_{aklb}(b,k \leq m) - \gamma^{(m)}_{al}\delta_{bk}(b,k \leq m) - {} \\
&\eqspace \Gamma^{(m)}_{aklb}(a,l \leq m) - \gamma^{(m)}_{kb}\delta_{al}(a,l \leq m),
\end{aligned}
\end{multline}
where we indicated between brackets after each term the conditions that need to be satisfied for the particular term to be present. To reduce the hessian to only the unique elements of $\mat{X}$, we will assume that $\mat{X}$ is real to keep the equations simple. The second order derivative with respect to the upper triangle of $\mat{X}$ at $\mat{X} = \mat{0}$ can now be constructed as
\begin{align}
\frac{\ud^2\Ncal}{\ud X_{kl} \ud X_{ab}}
&= \frac{\du^2\Ncal}{\du X_{kl}\du X_{ab}} - \frac{\du^2\Ncal}{\du X_{kl}\du X_{ba}} - {} \\*
&\eqspace
\frac{\du^2\Ncal}{\du X_{lk}\du X_{ab}} + \frac{\du^2\Ncal}{\du X_{lk}\du X_{ba}} \\
&= 2\bigl(\Gamma^{(m)}_{kabl} + \Gamma^{(m)}_{kbal} + \gamma^{(m)}_{bl}\delta_{ak} - \gamma^{(m)}_{ak}\delta_{bl}\bigr)
\end{align}
for $a,k \leq m < b,l$.

\bibliographystyle{model1a-num-names}
\bibliography{bible}

\section{Supplementary Material: Worst case for highest occupied NO guess}
The worst case for the highest occupied NO guess in our sample of $10\,000$ randomly generated full CI wave functions is very strongly correlated as can be seen from the occupation numbers reported in Table~\ref{tab:worstCaseOccs}. Such a large spread of the particles over the NOs is not very often encountered, though such large spreads have been reported for quantum Hall droplets~\cite{ToloHarju2010}.

\begin{table}[b!]
  \centering
  \caption{The NO occupation numbers for the worst case found for the highest occupied NO guess.}
  \label{tab:worstCaseOccs}
  \begin{tabular}{R e C R e}
    \toprule%
    k		&\multicolumn{1}{C}{n_k}	&\qquad\qquad	&k	&\multicolumn{1}{C}{n_k} \\
    \otoprule
    1		&0.510\,695		&&11		&0.238\,700	\\
    2		&0.346\,783		&&12		&0.113\,964	\\
    3		&0.331\,745		&&13		&0.072\,028	\\
    4		&0.329\,818		&&14		&0.061\,078	\\
    5		&0.319\,462		&&15		&0.052\,088	\\
    6		&0.306\,393		&&16		&0.049\,527	\\
    7		&0.301\,910		&&17		&0.048\,225	\\
    8		&0.289\,709		&&18		&0.046\,853	\\
    9		&0.258\,433		&&19		&0.040\,833	\\
    10		&0.250\,992		&&20		&0.030\,765	\\
    \bottomrule
  \end{tabular}
\end{table}

\begin{table}[t]
  \centering
  \caption{The contribution of determinants to the norm where the 4 highest occupied NOs are exclusively present. This data is for the worst case situation of the highest NO guess when eliminating 16 orbitals.}
  \label{tab:detContrib}
  \begin{tabular}{C R@{.}L C C R@{.}L}
    \toprule%
    \multicolumn{1}{c}{NOs} &\multicolumn{2}{c}{contribution} &\qquad\qquad
      &\multicolumn{1}{c}{NOs} &\multicolumn{2}{c}{contribution} \\
    \otoprule
    1,2,3,4		&4&4\cdot10^{-11}		&&1,2,3	&9&2\cdot10^{-5} \\
			&\multicolumn{2}{c}{}		&&1,2,4	&9&6\cdot10^{-4} \\
    1,2		&1&0\cdot10^{-2}		&&1,3,4	&8&8\cdot10^{-4} \\
    1,3		&3&7\cdot10^{-2}		&&2,3,4	&4&5\cdot10^{-3} \\
    1,4		&2&6\cdot10^{-1} \\
    2,3		&2&6\cdot10^{-1} \\
    2,4		&1&2\cdot10^{-3}		&&1		&2&1\cdot10^{-1} \\
    3,4		&2&2\cdot10^{-3}		&&2		&7&2\cdot10^{-2} \\
			&\multicolumn{2}{c}{}		&&3		&2&9\cdot10^{-2} \\
    0			&5&8\cdot10^{-2}		&&4		&6&4\cdot10^{-2} \\	
    \bottomrule
  \end{tabular}
\end{table}

We analyzed the composition of the wave function in NO basis by counting the total contribution, $\Ncal_{\coord{k}}$, of the determinants for each combination of the 4 highest occupied NOs, which are reported in Table~\ref{tab:detContrib}. Let us explain in more detail what we mean by this. The first number in Table~\ref{tab:detContrib} is the total contribution to the norm of determinants that are composed of the four highest occupied NOs: 1, 2, 3 and 4. There is only one such determinant for a 4 electron system, so $\Ncal_{1,2,3,4} = \abs{c_{1,2,3,4}}^2$. The entries with only three of the highest occupied NOs indicate the total contribution of the determinants composed of these three orbitals and \emph{not} the other one of the 4 highest occupied NOs. For example, the entry $1,2,3$ indicates the sum of the contribution of all determinants with the highest 3 occupied NOs, but not with NO number 4
\begin{align}
\Ncal_{1,2,3} =  \sum_{i \neq 4}\abs{c_{1,2,3,i}}^2 = \sum_{i}\abs{c_{1,2,3,i}}^2 - \Ncal_{1,2,3,4}.
\end{align}
The other entries are similar. For example, the pair $(2,4)$ means the total contribution of determinants with only NOs 2 and 4 present and not NOs 1 and 3, i.e.
\begin{align}
\Ncal_{2,4} &= \half\sum_{i,j\neq1,3}\abs{c_{2,4,i,j}}^2 \\
&= \half\sum_{i,j}\abs{c_{2,4,i,j}}^2 - \Ncal_{1,2,4} - \Ncal_{2,3,4} - \Ncal_{1,2,3,4}.
\end{align}
The entry 0 is a bit special notation. With the 0 we mean the total contribution to the norm of the determinants not containing any of the the four highest occupied NOs
\begin{align}
\Ncal_0 \isDefinedAs \frac{1}{4!}\sum_{i,j,k,l \neq 1,2,3,4}\abs{c_{i,j,k,l}}^2.
\end{align}
From Table~\ref{tab:detContrib} we immediately see that the contribution of the determinant containing all four highest occupied NOs makes a marginal contribution to the wave function. The determinants containing three of the 4 highest occupied NOs already make a more significant contribution, though still not very significant. The largest contribution seems to come from determinants that only contain 1 or 2 of the four highest occupied NOs. Especially the total contribution of determinants containing the pairs $(1,4)$ and $(2,3)$ give a very large contribution and also the determinants only containing only the highest occupied NO.

It is important to realize that the very low contribution by the determinant constructed from only the 4 highest occupied NOs is only possible when the other NOs also have a significant occupation, such that the $\Ncal_{i,j}$ and $\Ncal_i$ can become large and still produce the occupation number spectrum. Hence, these problematic wave functions for the NO guess could only occur for very strongly correlated systems, such as quantum Hall droplets~\cite{ToloHarju2010}.

\end{document}